\documentclass[aps,prd,twocolumn,superscriptaddress]{revtex4-2}
\usepackage{amsmath}

\usepackage{amsfonts}
\usepackage{amssymb}
\usepackage{mathtools}
\usepackage{dsfont}
\usepackage[mathscr]{eucal}
\usepackage{braket}

\newcommand{\rmd}{\ensuremath{\mathrm{d}}}
\newcommand{\rme}{\ensuremath{\mathrm{e}}}
\newcommand{\rmi}{\ensuremath{\mathrm{i}}}

\DeclareMathOperator{\sech}{sech}
\DeclareMathOperator{\csch}{csch}
\DeclareMathOperator{\arcsinh}{arcsinh}

\DeclareMathOperator{\Tr}{Tr}

\begin{document}

%\preprint{}

\title{Unruh effect for detectors in superposition of accelerations}



\author{Luis C.\ Barbado}
\email[]{luis.cortes.barbado@univie.ac.at}
\affiliation{Institute for Quantum Optics and Quantum Information, Austrian Academy of Sciences, Boltzmanngasse 3, 1090 Vienna, Austria}
\affiliation{Quantum Optics, Quantum Nanophysics and Quantum Information, Faculty of Physics, University of Vienna, Boltzmanngasse 5, 1090 Vienna, Austria}

\author{Esteban Castro-Ruiz}
\email[]{esteban.castro.ruiz@ulb.ac.be}
\affiliation{Institute for Quantum Optics and Quantum Information, Austrian Academy of Sciences, Boltzmanngasse 3, 1090 Vienna, Austria}
\affiliation{Quantum Optics, Quantum Nanophysics and Quantum Information, Faculty of Physics, University of Vienna, Boltzmanngasse 5, 1090 Vienna, Austria}
\affiliation{QuIC, Ecole Polytechnique de Bruxelles, C.P.~165, Universit\'e Libre de Bruxelles, 1050 Brussels, Belgium}

\author{Luca Apadula}
\email[]{luca.apadula@univie.ac.at}
\affiliation{Institute for Quantum Optics and Quantum Information, Austrian Academy of Sciences, Boltzmanngasse 3, 1090 Vienna, Austria}
\affiliation{Quantum Optics, Quantum Nanophysics and Quantum Information, Faculty of Physics, University of Vienna, Boltzmanngasse 5, 1090 Vienna, Austria}

\author{\v{C}aslav Brukner}
\email[]{caslav.brukner@univie.ac.at}
\affiliation{Institute for Quantum Optics and Quantum Information, Austrian Academy of Sciences, Boltzmanngasse 3, 1090 Vienna, Austria}
\affiliation{Quantum Optics, Quantum Nanophysics and Quantum Information, Faculty of Physics, University of Vienna, Boltzmanngasse 5, 1090 Vienna, Austria}


\date{\today}

\begin{abstract}
The Unruh effect is the phenomenon that accelerated observers detect particles even when inertial observers experience the vacuum state. In particular, uniformly accelerated observers are predicted to measure thermal radiation that is proportional to the acceleration. Here we consider the Unruh effect for a detector that follows a quantum superposition of different accelerated trajectories in Minkowski spacetime. More precisely, we analyse the excitations of a pointlike multilevel particle detector coupled to a massless real scalar field and moving in the superposition of accelerated trajectories. We find that the state of the detector excitations is, in general, not a mere (convex) mixture of the thermal spectrum characteristics of the Unruh effect for each trajectory with well-defined acceleration separately. Rather, for certain trajectories and excitation levels, and upon the measurement of the trajectory state, the state of the detector excitations features in addition off-diagonal terms. The off-diagonal terms of these ``superpositions of thermal states'' are related to the distinguishability of the different possible states in which the field is left after its interaction with detector's internal degrees of the freedom.
\end{abstract}

%\keywords{}

\maketitle

%-----------------------------------------------------------------------------------------------
\section{Introduction}
%-----------------------------------------------------------------------------------------------

The Unruh effect is one of the cornerstone results of Quantum Field Theory in non-inertial frames or curved spacetime (QFT-CS), and the paradigmatic example of the frame-dependent notion of the particle content of a field. In simple words, this effect consists of the perception of thermal bath of particles of a quantum field by observers following constant acceleration trajectories in Minkowski spacetime, when the field is in the Minkowski vacuum state (the vacuum state for inertial observers). The temperature of the bath is the Unruh temperature $T_{\mathrm{U}} := a \hbar / (2 \pi k_{\mathrm{B}} c)$, proportional to the acceleration of the trajectory~$a$, with~$\hbar$, $k_{\mathrm{B}}$ and~$c$ being the Planck constant, the Boltzmann constant and the speed of light, respectively. First proposed by Unruh in~\cite{PhysRevD.14.870} as an effect closely related to the celebrated black hole radiation proposed by Hawking~\cite{Hawking:1974sw}, the phenomenon has been widely studied in the literature from different perspectives. See, for example,~\cite{Birrell:1982ix} for an introductory approach, \cite{Wald:1995yp} for a mathematically rigorous derivation, and~\cite{RevModPhys.80.787} for an extensive review on the Unruh effect and its applications.

Out of the different aspects of the Unruh effect, arguably one of the most important is the fact that it provides a clear example on how the description of a quantum field changes in terms of particle content when the reference frame used to describe it changes. This non-trivial change in the description of the field is peculiar to QFT-CS. Notice that in this theory the background geometry and the reference frames in which the field is described are classical. In contrast, in non-relativistic Quantum Mechanics (QM) a change to a non-inertial classical reference frame at most introduces an effective gravitational potential in the dynamics under observation.

However, in QM one can introduce the notion of Quantum Reference Frame (QRF) and changes between QRFs as a generalization which allows to consider quantum mechanical systems as reference frames. In a recent work~\cite{Giacomini:2017zju} it has been shown that changes between QRFs in non-relativistic QM give rise to the frame dependence of the notions of quantum superposition and entanglement, and a generalization of the covariance principle. It seems natural then to attempt to extend the construction of QRFs to QFT-CS, and see whether further novel effects arise of the construction. Developing a fully consistent construction of a notion of QRFs for QFT-CS seems however a highly non-trivial task. On the road to it, we can nonetheless approach more concrete problems, that have their own interest and can help to shed some light on the topic.

In this work, we develop a description of the Unruh effect for observers which do not have a well-defined acceleration, but rather follow a superposition of trajectories with different accelerations in Minkowski spacetime. With this development we provide a first description of the Unruh effect in a particular family of QRFs, namely those which correspond to the superposition of accelerated trajectories which share the same Rindler wedge. We tackle this problem through the approach to the Unruh effect which makes use of particle detectors (see e.g.~\cite{PhysRevD.14.870, Birrell:1982ix, dewitt}). As our cornerstone result, we will find that the excitation of a particle detector interacting with the field and undergoing a superposition of accelerations is not always just an incoherent mixture of the thermal excitations that it would experience along each of the superposed trajectories individually. Rather, in addition to the mixture of planckian distributions for the different trajectories, coherent superpositions of excited states of the detector appear under certain conditions. In that sense, we can speak about ``superposition of thermal states''. These coherences appear because, under superposition of accelerations, the internal and external degrees of freedom of the detector can get entangled \emph{through} the interaction with the field, this entanglement remaining even after tracing out the field. Therefore, after measuring the external degrees of freedom in a certain basis, the state of the internal degrees of freedom can present coherences between different energy levels. This is in contrast to the usual role played by the Unruh effect as a source of decoherence for accelerated systems (see e.g.~\cite{Wang_2010, PARENTANI1995227, PhysRevX.9.011007}) \footnote{When considering the incidence of the Unruh effect on the entanglement between two \emph{different} detectors, the effect can lead both to destruction of entanglement (see e.g.~\cite{FuentesSchuller:2004xp,Alsing:2006cj,MartinMartinez:2010ds}) and to entanglement harvesting (see e.g.~\cite{MartinMartinez:2010ds,Salton:2014jaa,Koga:2019fqh})}.

The most common particle detector model for the study of the Unruh effect is the Unruh-DeWitt detector~\cite{dewitt}, which corresponds to a pointlike detector with two internal energy levels, weakly coupled to a real scalar field through a monopole interaction. As it will become clear in the article, this simplest model would not allow for the coherences that we mention before, and it is therefore not enough for our purposes. In this work we will consider a model of detector identical to the Unruh-DeWitt detector except for two major modifications: Our model contains three or more (eventually infinite) internal energy levels, and its trajectory is not a classically well-defined one, but rather can be a quantum superposition of different well-defined trajectories. The introduction of more than two internal energy levels already gives rise to the coherences we wish to describe. In particular, this model includes the use of a harmonic oscillator for the internal degree of freedom of the detector considered in~\cite{Lin:2006jw,MartinMartinez:2010sg,Dragan:2011hq,Brown:2012pw}.

The explicit consideration of the trajectory of the detector as a quantum degree of freedom, that can present an in principle arbitrary delocalization, is the real novelty of our approach, and what introduces a first step towards the description of the Unruh effect in a QRF. There are some previous works which consider coherent superpositions of trajectories of the detector, which however use different constructions and/or for different purposes. In~\cite{aleks2017simulating}, the authors consider a pair of Rindler observers in Minkowski background in a state of quantum superposition of having two different values of proper acceleration, their purpose being to realize indefinite causal order---a situation in which causal relations between events are subjected to quantum superposition---. In~\cite{Parentani:1995iw} a modeling of a particle detector using wave packets of two massive quantum fields with slightly different rest mass is considered, the localization of the wave packets being then a quantum degree of freedom; but only highly localized wave packets are considered, the purpose of introducing the external degree of freedom being to give account for the recoils on the detector produced by the emission of particles. In~\cite{MartinMartinez:2010sg} the authors describe an experiment involving the superposition of an Unruh-DeWitt detector along an inertial and an accelerated trajectory. The purpose would be to detect the difference in the Berry phase produced by the Unruh effect in the interference pattern of the two trajectories. However, this would be the only aim of the superposition of trajectories, which is otherwise not explicitly considered as a quantum degree of freedom. More recently, in~\cite{stritzelberger2019coherent} the authors consider explicitly a first quantization of the trajectory of the detector in a way similar to the one considered in this article, but staying in the non-relativistic regime and thus not giving account for the Unruh effect. Finally, in a very recent article~\cite{foo2020unruhdewitt} the authors consider an Unruh-DeWitt detector in a superposition of trajectories with a construction of such superposition analogous to the one considered here. However, their computations are focused on the excitation rate of the detector for different superpositions of trajectories and different switching functions, finding the usual (fully decohered) Unruh effect just as a particular side result. We shall also point out that the spatial quantum superposition of trajectories considered here is completely different to the use of finite-size detectors, as for example in~\cite{PhysRevD.14.870,Grove_1983,Louko:2006zv,CASADIO1999109,CASADIO199533}, which may consists of finite-size boxes or spatially smeared interactions with the field.

In describing the Unruh effect under superposition of accelerations, we also discuss the state in which the field is left after the detector got excited. This discussion arises naturally when addressing the physical reason for the coherences that we find in the detector. These coherences have their origin in the overlap between the states in which the field is left when the detector gets excited along two trajectories with different well-defined acceleration and to different well-defined internal levels. These states of the field are not always fully distinguishable, and therefore no full decoherence is introduced after tracing out the field in order to describe the state of the detector. In analyzing the overlap between those different states of the field, we will be able to further characterize them. A characterization of these states in the Minkowski reference frame can be found in~\cite{PhysRevD.38.1118}. Under the circumstances that we consider for the interaction (large interaction time with weak interaction, and superpositions of spatially localized trajectories), the states in which the field is left are those in which a Rindler particle from the thermal bath has been absorbed by the detector~\cite{PhysRevD.29.1047, PhysRevD.48.3731}, this absorption being almost fully delocalized in time, and therefore the absorbed particle having negligible dispersion in frequency; while being partially localized in space around the trajectory of the detector, which in the Rindler reference frame is a static trajectory. We find that a critical condition for the states in which the field is left, corresponding to different trajectories and different excitation levels of the detector, not to be fully distinguishable is that the energy of the absorbed particle is the same as measured by \emph{any} Rindler (accelerated) observer in the given Rindler wedge. This is in complete agreement with the fact that Minkowski vacuum state is indeed a thermal bath as perceived by any accelerated observer, the local temperature perceived by different observers being different simply because of the non-trivial Tolman factor~\cite{Tolman1934-TOLRTA} of the metric in the Rindler wedge.

The article is organized as follows. In Section~\ref{statement} we set up the problem, introducing the field, the detector model, the trajectories and the interaction. In Section~\ref{results} we give the results obtained for the state of the detector after the interaction, both for the full internal and external degrees of freedom, and for the internal degrees of freedom after measuring the external ones. We discuss physically the interpretation of the different results. In Section~\ref{example} we give an example of a detector following a superposition of trajectories, in which we can visualize the structure of coherences present for the internal degrees of freedom. Finally, in Section~\ref{discussion} we discuss possible extensions of the construction considered in this article. In Appendix~\ref{computation} we provide the detailed calculations yielding the results in Section~\ref{results}. In Appendix~\ref{function_f} we provide some further analytic expressions for the factor that determines the intensity of the coherences appearing in the state of the detector. In Appendix~\ref{continuous} we briefly consider the case in which the degrees of freedom of the detector (both the internal and the external) have a continuous spectrum.

%-----------------------------------------------------------------------------------------------
\section{Statement of the problem}\label{statement}
%-----------------------------------------------------------------------------------------------

Throughout the article we will consider natural units $\hbar = c = k_{\mathrm{B}} = 1$. Let us consider a real scalar massless quantum field~$\hat{\phi} (T,X,Y,Z)$ in Minkowski spacetime. Coupled to it, we consider a pointlike detector with several internal excitation levels~$\{\ket{0}_{\mathrm{D}}, \ket{\omega_1}_{\mathrm{D}}, \ket{\omega_2}_{\mathrm{D}},\ldots \}$, with energies~$0 < \omega_1 < \omega_2 < \ldots$ (there can be a finite or an infinite number of levels). The detector has also an external degree of freedom corresponding to the trajectory that it follows. We will consider trajectories with constant acceleration in the Rindler wedge~$Z > |T|$ (therefore, accelerated in the $Z$-direction towards increasing~$Z$). This wedge is covered by the Rindler coordinates~$(t,x,y,z)$, with~$z > 0$, related to the Minkowski coordinates~$(T,X,Y,Z)$ by
\begin{equation}
T = z \sinh(a t), \quad X = x, \quad Y = y, \quad Z = z \cosh(a t);
\label{rindler_coordinates}
\end{equation}
where~$a > 0$ is an arbitrary parameter with dimension of acceleration. The metric in these coordinates reads
\begin{equation}
\rmd s^2 = - (a z)^2 \rmd t^2 + \rmd x^2 + \rmd y^2 + \rmd z^2.
\label{metric}
\end{equation}

\begin{figure}
\includegraphics[width=\columnwidth]{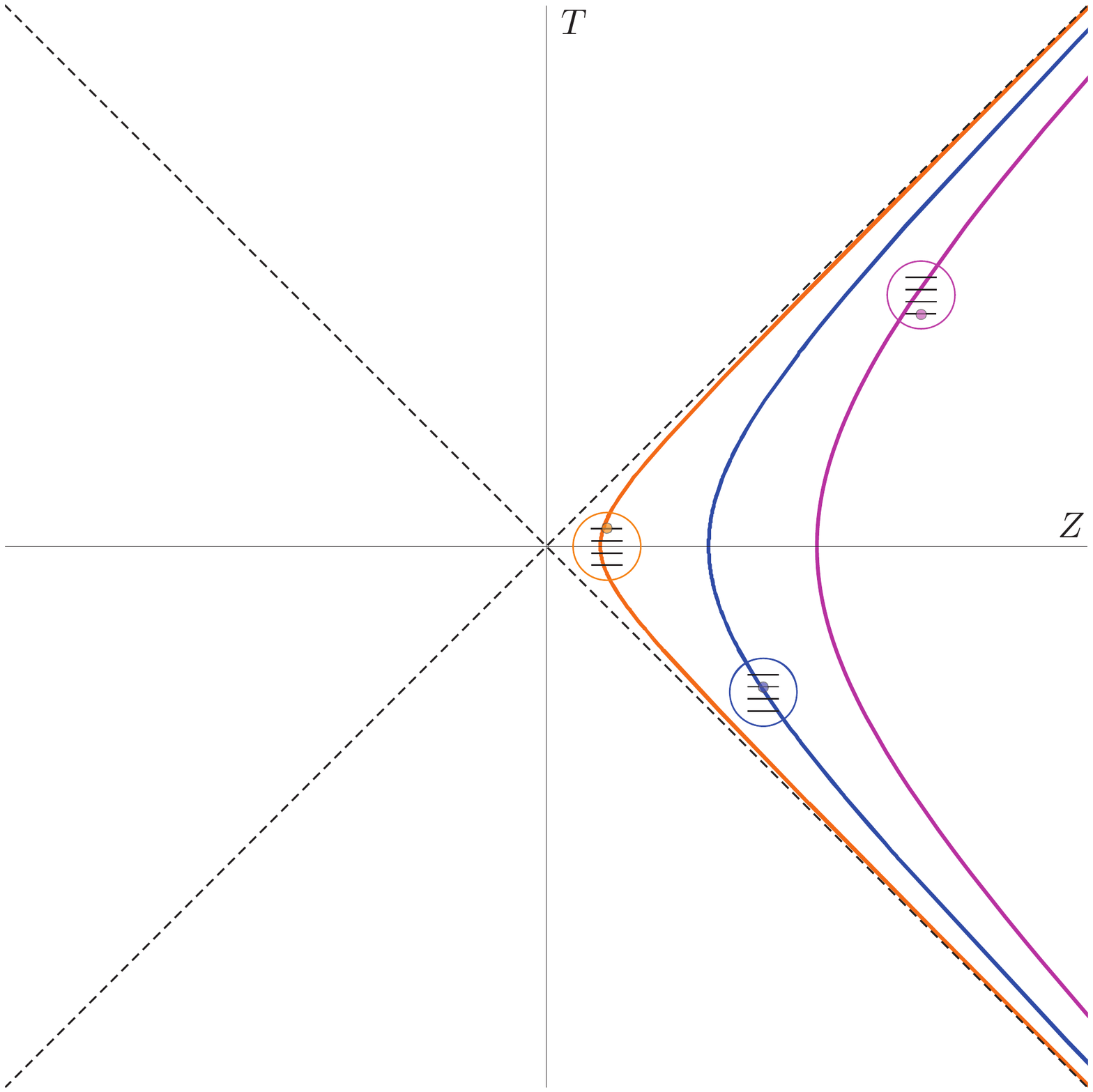}
\caption{\label{unruhplot} A multi-level Unruh-De Witt detector moves in a superposition of constant-acceleration trajectories. Each trajectory in the superposition is depicted by a hyperbola. The different colors mean that each hyperbola corresponds to a branch of the superposition. The detector interacts with a quantum field, getting excited in a way that depends on its state of motion. Upon measurement in a ``superposition of trajectories'' (see main text), the state of the detector can exhibit coherences between the different accelerations (corresponding to different temperatures) of the superposition.}
\end{figure}

The Hilbert space of the external degree of freedom of the detector (its trajectory) is spanned by the states~$\{\ket{1}_{\mathrm{T}}, \ket{2}_{\mathrm{T}}, \ldots \}$. For the states in this basis, the trajectory of constant acceleration is well-defined and given by
\begin{equation}
(\hat{t}(\tau), \hat{x}(\tau), \hat{y}(\tau), \hat{z}(\tau)) \ket{n}_{\mathrm{T}} = (\tau/(a z_n), x_n, y_n, z_n) \ket{n}_{\mathrm{T}},
\label{trajectory}
\end{equation}
where~$\tau$ is the proper time of the detector and~$x_n$, $y_n$, and~$z_n$ are constants. This corresponds to a semiclassical trajectory of constant acceleration~$a_n := 1/z_n$. We consider that all trajectories are fully distinguishable, so that~$\braket{n|m}_{\mathrm{T}} = \delta_{n m}$. For convenience, we organize them by increasing~$z_n$ (decreasing acceleration), $0 < z_1 \leq z_2 \leq \ldots$ (again, the different trajectories considered may be finite or infinite). We notice here that the time in which the unitary evolution of the system takes place is the detector's proper time~$\tau$, which stays as a parameter. The operator~$\hat{t}(\tau)$ corresponds to the Rindler coordinate time at which the detector is at some given~$\tau$ along a given trajectory, which will take different values for the different trajectories. In Figure~\ref{unruhplot} we plot an example of a superposition of these trajectories.

We will work in the interaction picture. The detector is coupled to the field with the following interaction term in the action:
\begin{equation}
\hat{S}_{\mathrm{I}} = \varepsilon \int_{-\infty}^{\infty} \rmd \tau \chi (\tau) \hat{m} (\tau) \hat{\phi} (\hat{x}(\tau)),
\label{coupling}
\end{equation}
where~$0 < \varepsilon \ll 1$ is a weak coupling constant, $\chi (\tau) \geq 0$ is a switching function that controls the intensity of the coupling in time, $\hat{x}(\tau)$ is the ``trajectory operator'', which action is given in~(\ref{trajectory}); and~$\hat{m} (\tau)$ is the monopole moment of the detector. We will work in first order perturbation theory in the coupling constant~$\varepsilon$.

We consider the switching function to be given by the square root of a Gaussian function with width~$T$, which is the approximate time duration of the interaction:
\begin{equation}
\chi (\tau) = \frac{1}{(2 \pi)^{1/4}} \rme^{-\tau^2 / (4 T^2)}.
\label{switching}
\end{equation}
When considering switching functions in the interaction, the switching on and off process can itself excite the detector, this effect getting scrambled with the excitations due to the Unruh effect in a way which does not always allow for a clear separation~\cite{Sriramkumar_1996,Satz_2007,Shevchenko_2017}. In order to avoid this situation, we need to consider smooth switching functions with an interaction time much larger than the inverse of the minimum frequency~$\omega_1$ that we wish to explore. As we will be able to check in the results, any effect due to the finite-time interaction becomes then negligible as compared to those due to the Unruh effect. On the other hand, too large interaction times can yield an excitation probability that goes beyond first order in~$\varepsilon$, breaking the validity of the results of the perturbative approach. As one can check following the computations in Appendix~\ref{computation}, a compromise value for the interaction time~$T$, which avoids the contribution of switching transients while keeping the consistency with first order perturbation, is the following:
\begin{equation}
T \sim \frac{1}{\varepsilon \omega_1} \gg \frac{1}{\omega_1} \geq \frac{1}{\omega_i}.
\label{duration}
\end{equation}

Too high accelerations may also yield an excitation probability beyond first order in~$\varepsilon$. Because of that, consistency with first order perturbation also requires the following limitation for the highest acceleration~$a_1 = 1/z_1$ (see the end of Appendix~\ref{computation}):
\begin{equation}
a_1 \lesssim \omega_1 / \mu, \quad \mu := \frac{1}{2 \pi} \log \left(\frac{1}{2\pi} + 1 \right) \simeq 0.02.
\label{acceleration_limitation}
\end{equation}
This limitation implies that $\omega_i \geq \omega_1 \gtrsim \mu a_1 \geq \mu a_n$, and therefore that the arbitrarily low frequency regime cannot in principle be explored. However, notice that the symbol~$\sim$ must be understood as a limitation in order of magnitude as compared to~$\varepsilon$; that is, we could have $\omega_i / (\mu a_n) < 1$ as far as the quotient remains significantly greater than~$\varepsilon$ [$\omega_i / (\mu a_n) \gg \varepsilon$]. Therefore, for arbitrarily weak coupling, one could expand the lower limit of the frequency range as desired.

The monopole moment evolving with the free Hamiltonian of the detector is given by
\begin{equation}
\hat{m} (\tau) = \sum_i \zeta_i\ \rme^{\rmi \omega_i \tau} \ket{\omega_i} \bra{0}_{\mathrm{D}} + \mathrm{h.c.};
\label{monopole}
\end{equation}
with~$\zeta_i$ being the coupling amplitudes from the ground state to the different excited states (we only consider coupling with the ground state since we only work in first order perturbation theory around this state). We impose that~$|\zeta_i| \lesssim 1$ to keep the interaction term of order~$\varepsilon$.

We consider the initial state of the system (detector and field) to be
\begin{equation}
\ket{\Psi (\tau \to -\infty)} = \ket{0}_{\mathrm{D}} \ket{0}_{\mathrm{F}} \left( \sum_n A_n \ket{n}_{\mathrm{T}} \right),
\label{initial_state}
\end{equation}
where~$\ket{0}_{\mathrm{F}}$ is the Minkowski vacuum state of the field, and~$A_n$ are the normalized amplitudes of the different accelerated trajectories of the detector. After the interaction has taken place, the state to first order in~$\varepsilon$ is
\begin{equation}
\ket{\Psi (\tau \to \infty)} \approx \left( \hat{\mathrm{I}} + \rmi \hat{S}_{\mathrm{I}} \right) \ket{\Psi (\tau = -\infty)}.
\label{final_state}
\end{equation}

%-----------------------------------------------------------------------------------------------
\section{Results}\label{results}
%-----------------------------------------------------------------------------------------------

%-----------------------------------------------------------------------------------------------
\subsection{State of the detector after the interaction}
%-----------------------------------------------------------------------------------------------

We need to compute the second term in~(\ref{final_state}), which will contain excited states of internal energy levels of the detector that will be different for each component of the acceleration. In most approaches to the computation of the excitation of the detector (see e.g.~\cite{Birrell:1982ix,Barbado:2012fy,foo2020unruhdewitt}) the tracing out of the field degrees of freedom is taken in the first place, yielding expressions for the excitation probabilities or rates in terms of two-point correlation functions of the field. In contrast, for the present work it is more convenient to compute the different states of the field explicitly, before taking the trace. In a generic way, we can write
\begin{align}
\ket{\Psi (\tau \to \infty)} \approx\ & \ket{0}_{\mathrm{D}} \ket{0}_{\mathrm{F}} \left( \sum_n A_n \ket{n}_{\mathrm{T}} \right)
\nonumber \\
& + \rmi \varepsilon \sum_{i, n} \zeta_i A_n \ket{\omega_i}_{\mathrm{D}} \ket{\omega_i, n}_{\mathrm{F}} \ket{n}_{\mathrm{T}},
\label{final_state_2}
\end{align}
where
\begin{align}
\ket{\omega_i, n}_{\mathrm{F}} :=\ & (\rmi \varepsilon \zeta_i A_n)^{-1} \bra{\omega_i}_{\mathrm{D}} \bra{n}_{\mathrm{T}} \ket{\Psi (\tau \to \infty)}
\nonumber \\
=\ & (\varepsilon \zeta_i)^{-1} \bra{\omega_i}_{\mathrm{D}} \bra{n}_{\mathrm{T}} \hat{S}_{\mathrm{I}} \ket{0}_{\mathrm{D}} \ket{0}_{\mathrm{F}} \ket{n}_{\mathrm{T}}
\label{state_field}
\end{align}
is the (not normalized) state in which the field is left for the trajectory~$\ket{n}_{\mathrm{T}}$ and the excited state~$\ket{\omega_i}_{\mathrm{D}}$ of the detector. In~(\ref{state_field}) we have used the fact that the trajectory operator contained in the action is diagonal in the basis of well-defined trajectories [see~(\ref{trajectory})]. If we now trace out the field in the final state in~(\ref{final_state_2}), we obtain generically the final state of the detector:
\begin{widetext}
\begin{multline}
\rho_{\mathrm{DT}} := \Tr_{\mathrm{F}} (\ket{\Psi (\tau \to \infty)}\bra{\Psi (\tau \to \infty)}) \approx \\
\ket{0}\bra{0}_{\mathrm{D}} \left( \sum_{n, m} A_n^* A_m \ket{m} \bra{n}_{\mathrm{T}} \right) + \varepsilon^2 \sum_{i, j, n, m} \zeta_i^* A_n^* \zeta_j A_m \braket{\omega_i, n|\omega_j, m}_{\mathrm{F}} \ket{\omega_j}\bra{\omega_i}_{\mathrm{D}}\ket{m} \bra{n}_{\mathrm{T}}.
\label{state_detector_gen}
\end{multline}

The quantities that remain to be calculated are the scalar products~$\braket{\omega_i, a_n|\omega_j, a_m}_{\mathrm{F}}$. These quantities are computed in detail in Appendix~\ref{computation}, under approximations consistent with the first order perturbation theory, including the large time approximation in~(\ref{duration}). The state of the detector after the interaction is finally given by
\begin{multline}
\rho_{\mathrm{DT}} \approx \ket{0}\bra{0}_{\mathrm{D}} \left( \sum_{n, m} A_n^* A_m \ket{m} \bra{n}_{\mathrm{T}} \right) + \frac{\varepsilon^2 T}{2 \pi} \Bigg[ \sum_{m} |A_m|^2 \ket{m}\bra{m}_{\mathrm{T}} \sum_{j} |\zeta_j|^2 \frac{\omega_j}{\rme^{2 \pi q_{j m}} - 1} \ket{\omega_j}\bra{\omega_j}_{\mathrm{D}} \\
+ \sum_{\mathclap{\substack{n, m \\ n \neq m}}} A_n^* A_m \ket{m}\bra{n}_{\mathrm{T}} \sum_{\mathclap{\substack{i, j \\ i \neq j}}}^{\mathrm{cond}} \zeta_i^* \zeta_j\ \Lambda^{ij}_{nm}\ \frac{\sqrt{\omega_i \omega_j}}{\rme^{2 \pi q_{j m}} - 1} \ket{\omega_j}\bra{\omega_i}_{\mathrm{D}} \Bigg];
\label{state_detector}
\end{multline}
\end{widetext}
where
\begin{equation}
q_{j m} := \omega_j z_m = \frac{\omega_j}{a_m},
\label{q_def}
\end{equation}
the label~`$\mathrm{cond}$' in the sum means that only the terms for which the condition
\begin{equation}
q_{i n} \approx q_{j m}
\label{quotients_q}
\end{equation}
holds to order~$\varepsilon$ are considered; and
\begin{equation}
\Lambda^{ij}_{nm} := \frac{\braket{\omega_i, n|\omega_j, m}_{\mathrm{F}}}{\sqrt{\braket{\omega_i, n|\omega_i, n}_{\mathrm{F}}\braket{\omega_j, m|\omega_j, m}_{\mathrm{F}}}}
\label{lambda_def}
\end{equation}
is the scalar product between the \emph{normalized} states of the field left for the trajectory~$\ket{n}_{\mathrm{T}}$ and the excited state~$\ket{\omega_i}_{\mathrm{D}}$, and for the trajectory~$\ket{m}_{\mathrm{T}}$ and the excited state~$\ket{\omega_j}_{\mathrm{D}}$. As shown in Appendix~\ref{computation}, when~(\ref{quotients_q}) is satisfied this normalized scalar product is given by a function of~$q_{j m}~(\approx q_{i n})$ in~(\ref{q_def}) and the relative quantities between the trajectories
\begin{align}
\Delta \xi_{m n} & := \log \left( \frac{z_m}{z_n} \right),
\nonumber \\
\Delta \bar{x}_{m n} & := \Delta x^\perp_{m n} \sqrt{\frac{1/z_m^2 + 1/z_n^2}{2}},
\label{relative_values} \\
\Delta x^\perp_{m n} & := |\vec{x}^\perp_m - \vec{x}^\perp_n|, \quad \vec{x}^\perp_m := (x_m, y_m);
\nonumber
\end{align}
as
\begin{equation}
\Lambda^{ij}_{nm} = \sqrt{\frac{\pi}{2} \sech \Delta \xi_{m n}}\ \frac{P^{-1/2}_{\rmi q_{j m} - 1/2} (u(\Delta \xi_{m n}, \Delta \bar{x}_{m n}))}{[u(\Delta \xi_{m n}, \Delta \bar{x}_{m n})^2 - 1]^{1/4}},
\label{f_weight}
\end{equation}
with
\begin{equation}
u(\Delta \xi_{m n}, \Delta \bar{x}_{m n}) := \cosh \Delta \xi_{m n} + \frac{\Delta \bar{x}_{m n}^2}{2} \sech \Delta \xi_{m n}
\label{u_distance}
\end{equation}
and $P^\mu_\nu (x)$ being the associated Legendre function of the first kind \footnote{For the arguments considered, this function is real and belongs to the class known as \emph{conical} or \emph{Mehler functions,} discussed in~\cite{NIST:DLMF}.}. We plot the value of~$\Lambda^{ij}_{nm}$ as a function of~$(\Delta \xi_{m n}, \Delta \bar{x}_{m n})$ for different values of~$q_{j m}$ in Figure~\ref{graphics}. In Appendix~\ref{function_f} we provide simpler formulas and graphs for the cases~$\Delta \xi_{m n} = 0$ and~$\Delta \bar{x}_{m n} = 0$.
\begin{figure*}[t]
\includegraphics[width=\textwidth]{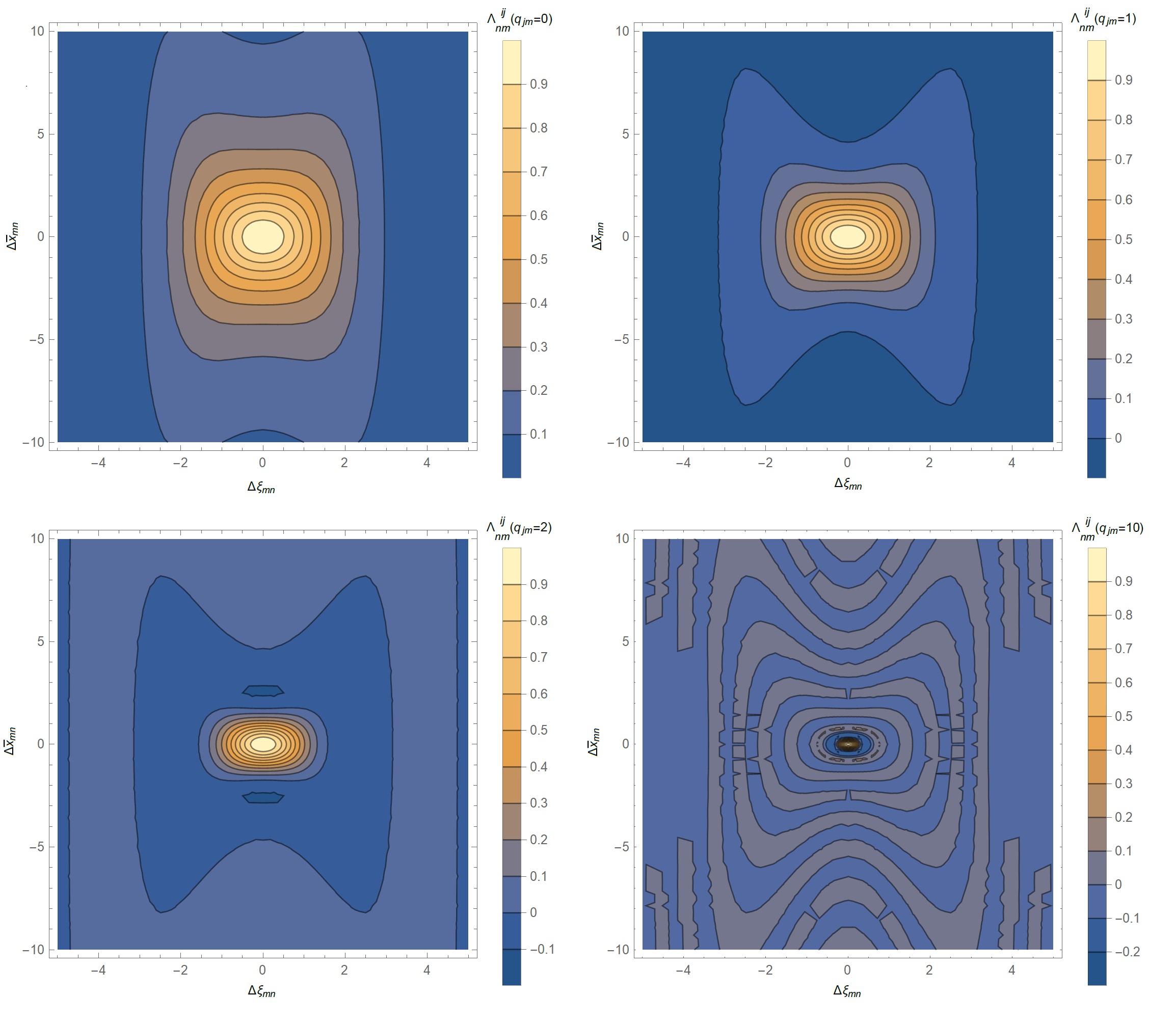}
\caption{\label{graphics} Scalar product~$\Lambda^{ij}_{nm}$ as a function of~$(\Delta \xi_{m n}, \Delta \bar{x}_{m n})$ for~$q_{j m} = 0$, $q_{j m} = 1$, $q_{j m} = 2$ and $q_{j m} = 10$.}
\end{figure*}

Two important remarks about~(\ref{state_detector}) are in order. First, even if the perturbative terms appear with a factor~$\varepsilon^2$, this does not mean that the terms are always of order~$\varepsilon^2$; since one has to take into account the factor~$T$ and its relative order of magnitude given by~(\ref{duration}), which also involves~$\varepsilon$. At the end of Appendix~\ref{computation} it is proven that the perturbative terms remain of order~$\varepsilon$ or smaller. Second, although the total time of interaction~$T$ still appears in~(\ref{state_detector}), it is just a multiplicative factor, and its appearance does not mean that the finite time considered introduces spurious contributions. The physically relevant large-time limit has already been taken by using~(\ref{duration}), and no contributions due to the transients remain in the expressions. Indeed, one can make~$T$ arbitrarily large by taking~$\varepsilon$ arbitrarily small (more interaction time with weaker coupling), and all the results obtained remain formally identical, while all approximations taken still hold, in fact more accurately.

%-----------------------------------------------------------------------------------------------
\subsection{Physical interpretation}
%-----------------------------------------------------------------------------------------------

Let us comment on the different terms appearing in~(\ref{state_detector}). First of all, we highlight that the parameter~$a$ with which we constructed the Rindler coordinates~(\ref{rindler_coordinates}) does not appear in any of the terms, as it should happen since it was just an auxiliary parameter with no physical meaning in the construction. The first term in~(\ref{state_detector}) is the contribution to zeroth order in~$\varepsilon$, and corresponds to the case in which the detector does not interact with the field. The terms with the factor~$\varepsilon^2 T$ correspond to the contribution of the interaction with the field. There are both diagonal and off-diagonal terms. The diagonal terms for each trajectory~$m$ follow a planckian probability distribution with the Unruh temperature~$a_m /(2 \pi)$, simply filtered by the coupling amplitudes~$\zeta_i$ for each frequency. These are the contributions of the Unruh effect for each trajectory separately, combined in an incoherent way. Therefore, our construction reproduces the canonical Unruh effect for quantum detectors as a particular case: A detector following a well-defined classical trajectory with constant acceleration~$a_m$ would get excited as if immersed in a thermal bath with temperature~$a_m / (2 \pi)$.

The novel result are the off-diagonal terms in the second line of~(\ref{state_detector}), corresponding to coherences between different trajectories. These terms only appear between trajectories and excited states for which the condition~(\ref{quotients_q}) holds. Physically, this condition entails that the quotients~$q_{j m}$ and~$q_{i n}$ of the frequencies~$\omega_j$ and~$\omega_i$ being excited along each trajectory~$\ket{m}_{\mathrm{T}}$ and~$\ket{n}_{\mathrm{T}}$, with the Tolman factor~\cite{Tolman1934-TOLRTA} along the corresponding trajectory~[$1/(a z_m)$ and~$1/(a z_n$), see the metric~(\ref{rindler_coordinates})], are (approximately) the same in both trajectories compared. Taking into account the role of the Tolman factor, this condition means that the two excited states of the detector must be degenerate in energy as described by any Rindler observer. We will comment in short why the coherences appear only when such condition is fulfilled. When~(\ref{quotients_q}) is satisfied, the corresponding off-diagonal term is the product of the square root of the planckian spectra for the two corresponding trajectories and excited frequencies \footnote{When~(\ref{quotients_q}) holds, it can be used consistently as an equality in the expressions, as we have done for the corresponding off-diagonal terms.}, weighted with the scalar product~$\Lambda^{ij}_{nm}$ given in~(\ref{f_weight}).

As we already advanced, the origin of the coherences found can be traced back to the properties of the state of the field left any time the detector gets excited~$\ket{\omega_i, n}_{\mathrm{F}}$, as the presence of the scalar product between states of the field~$\Lambda^{ij}_{nm}$ in the off-diagonal terms clearly shows. The perturbations left on the field corresponding to transitions to different energy levels of the detector and through different trajectories are not always distinguishable, but rather may overlap. When this is the case, the scalar product~$\Lambda^{ij}_{nm}$ is non-zero and the off-diagonal terms appear. This happens because, when the compared field states are not fully distinguishable, no full entanglement is created between the excited states of the detector and the field due to the interaction. Therefore, tracing out the field does not introduce full decoherence in the state of the detector.

Let us discuss now in more detail the non-distinguishability of the states of the field, depending on the trajectories and excited states of the detector to which the compared states of the field correspond. The non-distinguishability is clearly encoded in the condition~(\ref{quotients_q}) and the properties of the scalar product~$\Lambda^{ij}_{nm}$. Before discussing the functional dependence of~$\Lambda^{ij}_{nm}$ in~(\ref{f_weight}), let us bring out in the first place the geometric significance of the two quantities it depends on (beyond the already described quotient~$q_{j m}$), namely $\Delta \xi_{m n}$ and~$\Delta \bar{x}_{m n}$. $\Delta \xi_{m n}/a$ is the difference between the so-called \emph{Lass coordinate} $\xi_{\mathrm{Lass}} := \log (a z) /a$ of the two trajectories~\cite{doi:10.1119/1.1969430}, which is proportional to the radar distance \footnote{The radar distance between two trajectories is determined by half of the proper time, as measured by an observer following one of the trajectories, that runs between he sends a light ray, this ray is ``reflected'' in the other trajectory, and he crosses the reflected light ray again. Such quantity is kept constant between any two constant accelerated trajectories in the same Rindler wedge. See e.g.~\cite{Pauri:2000cr,Dolby:2001aa,Tilbrook:1997:GCF:269671.269672}.} between two accelerated observers with the same coordinates~$(x,y)$. Also, from the metric~(\ref{rindler_coordinates}) we can see that~$\Delta \bar{x}_{m n} z_m$ is the radar distance in any direction perpendicular to the acceleration for two trajectories for which~$z_n = z_m$. Therefore, the quantities~$\Delta \xi_{m n}$ and~$\Delta \bar{x}_{m n}$ provide a notion of (normalized) distance in the respective directions (parallel and perpendicular to the acceleration) in the Rindler reference frame.

Let us now describe the functional dependence of the scalar product~$\Lambda^{ij}_{nm}$ in~(\ref{f_weight}) (plotted in Figure~\ref{graphics}). It reaches its maximum value of~$1$ only for~$\Delta \xi_{m n} = \Delta \bar{x}_{m n} = 0$; that is, only if the compared trajectories coincide, which because of condition~(\ref{quotients_q}) means that the excited states of the detector also coincide. In this case, the term would simply not be an off-diagonal term, but rather a diagonal one. The functional dependence is even both in~$\Delta \xi_{m n}$ and in~$\Delta \bar{x}_{m n}$, and it decays to zero for large values of~$|\Delta \xi_{m n}|$ or~$|\Delta \bar{x}_{m n}|$, with oscillations around~$\Lambda^{ij}_{nm} = 0$ that become relatively more significant for higher~$q_{j m}$. The size of the region in the arguments~$(\Delta \xi_{m n}, \Delta \bar{x}_{m n})$ for which the function takes non-negligible values scales approximately as~$\sim 1/q_{j m}$ except for low values of~$q_{j m}$, for which it approaches a finite maximum size. Taking into account the geometric meaning of the arguments described before, the results obtained indicate that the coherence between trajectories decays for distant trajectories in the Rindler reference frame, the decay being sharper for higher frequencies.

Given the above discussion on the off-diagonal terms, we are in condition to give a clear physical picture for the appearance of the coherences that we find. In the Rindler reference frame, the excitation of the detector happens because it absorbs a particle of the thermal bath that it perceives~\cite{PhysRevD.29.1047, PhysRevD.48.3731}. We can interpret better the presence of both the condition~(\ref{quotients_q}) and the scalar product~$\Lambda^{ij}_{nm}$ for the off-diagonal terms in light of this picture of the interaction. On the one hand, the absorption of the particle is almost fully delocalized in time along all the interaction period, and therefore the absorbed particle in the Rindler frame has very little dispersion in frequency. This explains the necessity of the fine tuning of the frequencies required in~(\ref{quotients_q}): If the condition is not satisfied, the particles absorbed along the different trajectories would have fully distinguishable energies as described by any Rindler observer, the states of the field left would be fully distinguishable, and no off-diagonal terms would be left after tracing out the field. On the other hand, we can interpret the dependence of the scalar product~$\Lambda^{ij}_{nm}$ in the arguments~$(\Delta \xi_{m n}, \Delta \bar{x}_{m n})$ as providing a notion of spatial localization of the particle absorbed, which would be neither fully localized nor fully delocalized. Since two trajectories can have some finite non-negligible distance in the Rindler reference frame and still the effect of an absorption along each of them on the field be not fully distinguishable, we can conclude that the absorbed particles are delocalized ``in the surroundings'' of each trajectory, these surroundings having the shape as in Figure~\ref{graphics} in the Rindler reference frame in the transformed distances $\Delta \xi_{m n}$ and~$\Delta \bar{x}_{m n}$. We can identify the average size of the delocalization with a sort of wavelength of the absorbed particle in the corresponding coordinates. This wavelength is always finite, although it becomes arbitrarily small for high frequencies. However, as we already mentioned for arbitrarily low frequencies it does not become arbitrarily large, but rather reaches a maximum size, which order of magnitude (in the usual Rindler coordinates) is determined by the inverse of the accelerations involved.

A remarkable fact of our approach is that the scalar product between the states of the field~$\Lambda^{ij}_{nm}$ is computed using their representation in the Fock quantization associated to the Minkowski modes, in which the excitation of the detector is accompanied by the emission of a particle (see Appendix~\ref{computation}). However, the results have a much clearer physical interpretation in the Rindler reference frame. This further supports that the description of the effect in the two reference frames is complementary and yields no contradictions~\cite{RevModPhys.80.787, PhysRevD.29.1047, PhysRevD.48.3731}; as far as the different quantities (in particular the interaction time) and states involved are kept finite and normalizable, respectively.

%-----------------------------------------------------------------------------------------------
\subsection{State of the internal energy levels}
%-----------------------------------------------------------------------------------------------

We would like to have also a description of the state of the internal energy levels of the detector alone. If we trace the state in~(\ref{state_detector}) for the degree of freedom of the trajectory, we obtain
\begin{multline}
\rho^{\mathrm{Tr}}_{\mathrm{D}} := \Tr_{\mathrm{T}} (\rho_{\mathrm{DT}}) \approx \ket{0}\bra{0}_{\mathrm{D}} \\
+ \frac{\varepsilon^2 T}{2 \pi} \sum_n |A_n|^2 \left( \sum_i |\zeta_i|^2 \frac{\omega_i}{\rme^{2 \pi \omega_i/a_n} - 1} \ket{\omega_i}\bra{\omega_i}_{\mathrm{D}} \right).
\label{internal_trace}
\end{multline}
The detector has some probability to get excited given by a weighted mixture of thermal states (filtered by the coupling amplitudes~$\zeta_i$ for each frequency) with different temperatures proportional to the corresponding accelerations. Since we have assumed all the trajectories to be fully distinguishable, this is again consistent with the standard result on the Unruh effect (for well-defined trajectories) for the particle detector that we have considered.

We can also consider the state of the internal energy levels left when measuring the trajectory in some complementary basis and finding it to be e.g.\ $\ket{\eta}_{\mathrm{T}} := \sum_n B_n \ket{n}_{\mathrm{T}}$. Such state is (without normalization)
\begin{widetext}
\begin{align}
\rho^{\mathrm{measure}}_{\mathrm{D}} :=\ & \Tr_{\mathrm{T}} (\ket{\eta}\bra{\eta}_{\mathrm{T}} \rho_{\mathrm{DT}}) \approx \left( \sum_{n,m} B_m^* A_n^* B_n A_m \right) \ket{0}\bra{0}_{\mathrm{D}}
\nonumber \\
& + \frac{\varepsilon^2 T}{2 \pi} \sum_m \left\{ |B_m|^2 |A_m|^2 \left( \sum_j |\zeta_j|^2 \frac{\omega_j}{\rme^{2 \pi q_{jm}} - 1} \ket{\omega_j}\bra{\omega_j}_{\mathrm{D}} \right) \right.
\nonumber \\
& \left. + \sum_{n \neq m} B_m^* A_n^* B_n A_m \left[ \sum_{i,j}^{\mathrm{cond}} \zeta_i^* \zeta_j \Lambda^{ij}_{nm} \frac{\sqrt{\omega_i \omega_j}}{\rme^{2 \pi q_{j m}} - 1} \ket{\omega_j}\bra{\omega_i}_{\mathrm{D}} \right] \right\},
\label{internal_measure}
\end{align}
\end{widetext}

This is the main result of our work. We notice that the diagonal terms corresponding to the thermal contribution remain as in~(\ref{internal_trace}), just with different weights. Added to this thermal contribution, some off-diagonal terms appear. Therefore in general the internal state of the detector is not just mixture of states with well-defined energy.

Notice also that, if the state~$\ket{\eta}_{\mathrm{T}}$ is taken to be orthogonal to the initial trajectory state, then the coefficient of the element~$\ket{0}\bra{0}_{\mathrm{D}}$ vanishes. This means that the detector could be found in a trajectory orthogonal to the initial given only if it got excited and the trajectory got entangled with the internal levels \emph{through} the field (this entanglement remaining even after tracing out the field). Notice also that this entanglement between the trajectory and the internal levels can only happen if the different trajectories have different values of~$z_n$ (different accelerations), since otherwise the same excitations appear along all the trajectories. The fact that the trajectory state can be found to be orthogonal to the initial one after the interaction exemplifies that, within the construction that we consider, the trajectory is not simply a fixed constrain of the problem. Rather, the trajectory is truly a quantum degree of freedom subject to the interaction with the field, which will actually be modified by this interaction unless the initial state is a well-defined trajectory (since in the basis of well-defined trajectories the interaction term is diagonal acting over the trajectory degree of freedom).

%-----------------------------------------------------------------------------------------------
\section{An explicit example}\label{example}
%-----------------------------------------------------------------------------------------------

Let us consider the simple case of a detector which internal energy levels correspond to those of a harmonic oscillator. We normalize the dimensions by fixing the frequencies to~$\omega_i = i$. For simplicity, we consider that the detector does not discriminate frequencies in the coupling, so we take~$\zeta_i = 1$. We prepare the detector in a superposition of three accelerated trajectories at $x_n = y_n = 0$ for all~$n$ (they are not perpendicularly displaced with respect to each other) and $z_1 = 0.5$, $z_2 = 1$ and $z_3 = 1.5$ ($a_1 = 2$, $a_2 = 1$ and $a_3 = 2/3$) in the following way:
\begin{equation}
\ket{\Psi (\tau \to -\infty)} = \frac{1}{\sqrt{3}} \ket{0}_{\mathrm{D}} \ket{0}_{\mathrm{F}} ( \ket{1}_{\mathrm{T}} + \ket{2}_{\mathrm{T}} + \ket{3}_{\mathrm{T}} ).
\label{initial_example}
\end{equation}

After the interaction, we measure the trajectory in some basis containing the initial state of the trajectory, and consider the case in which we find it to be in such state; that is, we consider $B_n = A_n = 1/\sqrt{3}$ for all~$n$ in~(\ref{internal_measure}).

Writing down the explicit numerical results obtained is of no particular interest, but rather showing the structure of the non-vanishing matrix terms in~(\ref{internal_measure}) and their order of magnitude is. We provide below a matrix which elements are minus the logarithm of the absolute value of the elements in~(\ref{internal_measure}) divided by~$\varepsilon^2 T$, for the first twelve excited states:
\begin{multline}
\left( -\log_{10} (|(\rho^{\mathrm{measure}}_{\mathrm{D}})_{j i}|/(\varepsilon^2 T)) \right)_{1 \leq j, i \leq 12} \simeq \\
\left( \begin{array}{cccccccccccc}
3.1 & 4.4 & 6.0 & & & & & & & & & \\
4.4 & 4.2 & 9.7 & 7.0 & & 11 & & & & & & \\
6.0 & 9.7 & 5.4 & & & 9.8 & & & 14 & & & \\
& 7.0 & & 6.6 & & 18 & & 13 & & & & 19 \\
& & & & 7.9 & & & & & 16 & & \\
& 11 & 9.8 & 18 & & 9.2 & & & 26 & & & 18 \\
& & & & & & 10 & & & & & \\
& & & 13 & & & & 12 & & & & 34 \\
& & 14 & & & 26 & & & 13 & & & \\
& & & & 16 & & & & & 14 & & \\
& & & & & & & & & & 16 & \\
& & & 19 & & 18 & & 34 & & & & 17
\end{array} \right).
\label{matrix_example}
\end{multline}
Notice that the greater the entry the exponentially smaller the element in~(\ref{internal_measure}) in absolute value. The entries corresponding to~$-\log_{10} 0$ have been omitted. We can visualize the structure of seven ``alignments'' of the non-vanishing elements, each one with a different slope. Along each of them the ratio of the frequencies is $3$, $2$, $3/2$, $1$ (the diagonal), $2/3$, $1/2$ and~$1/3$; which are the possible ratios between the values of~$z_n$ (or $a_n$) along the different trajectories. These are the elements for which condition~(\ref{quotients_q}) is fulfilled for at least one pair of trajectories (in this case just one pair, except for the diagonal elements). Different choices of trajectories or energy levels would of course yield different structures of the non-vanishing elements, the only fact in common being the presence of the diagonal elements. We can also check that the diagonal elements are always greater than any other in the same row or column. This is due to two facts: First, in the diagonal elements all the trajectories contribute; and second, the contributions are not lowered by the scalar product~$\Lambda^{ij}_{nm}$.

%-----------------------------------------------------------------------------------------------
\section{Further discussion}\label{discussion}
%-----------------------------------------------------------------------------------------------

We have studied the excitation of a particle detector following a quantum superposition of semiclassical trajectories with well-defined acceleration due to the Unruh effect. When the trajectories under superposition all belong to the same Rindler wedge, we have found that the state of the internal degrees of freedom of the detector after the interaction with the field, upon measurement of the external degree of freedom in some complementary basis, can present coherent superpositions of different energy levels. Although we did not consider the superposition of trajectories which do not share the same Rindler wedge, out of the discussion on the origin of the coherences found in this article, we can argue that these coherences will not be present when the Rindler wedges differ significantly (this significance being arguably determined by the parameter~$\varepsilon$). The reason is that a static trajectory in some Rindler wedge is not static in any other Rindler wedge, and therefore its distance (in Rindler coordinates) with respect to a static trajectory in the second wedge will change in time. Being the time of interaction needed to properly give account for the Unruh effect significantly large (in the perturbative regime), any two static trajectories in two different Rindler wedges will be most of the time very separated from one another, as measured from any of the two wedges. But the origin of the coherences found is the overlap of the perturbations on the field for trajectories which remain at some distance for which the scalar product~$\Lambda^{ij}_{nm}$ is non-negligible. Therefore this overlap, and hence the coherences, will not be significant when the trajectories are most of the time very separated.

We can argue that observers following trajectories with different acceleration in the same Rindler wedge describe the spacetime surrounding their trajectory with a different metric. Therefore, in the spirit of the notion of QRFs, an observer following a quantum superposition of these trajectories would perceive a sort of ``quantum superposition of metrics'' of the spacetime. Considering the equivalence principle, we can relate this situation with that of an observer in a quantum superposition of different distances from a black hole, and its perception of the corresponding Hawking radiation; or even with the situation of an observer which feels the gravitational field of a source which is in a quantum superposition of different masses (and therefore producing a quantum superposition of metrics, see for example~\cite{castroruiz2019time}). Approaching these situations with the construction developed in this article will be the aim of future works by the authors.

\appendix

%-----------------------------------------------------------------------------------------------
\section{Computation of the scalar products of the states of the field}\label{computation}
%-----------------------------------------------------------------------------------------------

In this Appendix we go in detail through the computation of the scalar products~$\braket{\omega_i, a_n|\omega_j, a_m}_{\mathrm{F}}$. We do it in two steps: Computation of the states of the field~$\ket{\omega_i, n}_{\mathrm{F}}$ in the Fock quantization associated to Minkowski modes, and computation of the scalar products themselves in the large time regime given by~(\ref{duration}).

Let us compute the states of the field~$\ket{\omega_i, n}_{\mathrm{F}}$ defined in~(\ref{state_field}) in the first place. The field operator present in the interaction~(\ref{coupling}) evolves according to its free Hamiltonian. Since all the trajectories that we consider are constrained to the right Rindler wedge~$Z > |T|$, we can expand the field in that region using Rindler modes. This expansion is
\begin{multline}
\hat{\phi} (t, x, y, z) = \int_0^\infty \rmd \omega \int \rmd^2 \vec{k}_\perp [ \hat{a}^{\mathrm{R}}_{\omega \vec{k}_\perp} v_{\omega \vec{k}_\perp} (t, x, y, z) \\
+ (\hat{a}^{\mathrm{R}}_{\omega \vec{k}_\perp})^\dagger v_{\omega \vec{k}_\perp} (t, x, y, z)^* ],
\label{field}
\end{multline}
where~$v_{\omega \vec{k}_\perp} (t, x, y, z)$ are the Rindler modes defined in that wedge, given by~\cite{RevModPhys.80.787}
\begin{equation}
v_{\omega \vec{k}_\perp} (t, x, y, z) = \sqrt{\frac{\sinh(\pi \omega / a)}{4 \pi^4 a}} K_{\rmi \omega / a} ( k_\perp z ) \rme^{\rmi(\vec{k}_\perp \cdot \vec{x}^\perp-\omega t)},
\label{rindler_modes}
\end{equation}
with~$\vec{k}_\perp := (k_x, k_y)$, $k_\perp := |\vec{k}_\perp|$, $\vec{x}^\perp := (x, y)$ and~$K_\nu (x)$ the modified Bessel function of the second kind; and $\hat{a}^{\mathrm{R}}_{\omega \vec{k}_\perp}$, $(\hat{a}^{\mathrm{R}}_{\omega \vec{k}_\perp})^\dagger$ are the associated annihilation and creation operators.

Plugging the expansion of the field~(\ref{field}), the trajectory~(\ref{trajectory}) and the evolution of the monopole~(\ref{monopole}) in~(\ref{state_field}), while for convenience not replacing yet the explicit expression for the switching function~$\chi (\tau)$, we obtain
\begin{align}
\ket{\omega_i, n}_{\mathrm{F}} =\ & \int_{-\infty}^\infty \rmd \tau\ \rme^{\rmi \omega_i \tau} \chi (\tau) \int_0^\infty \rmd \omega \int \rmd^2 \vec{k}_\perp
\nonumber \\
& \times [ \hat{a}^{\mathrm{R}}_{\omega \vec{k}_\perp} v_{\omega \vec{k}_\perp} (\tau/(a z_n), x_n, y_n, z_n)
\nonumber \\
& + (\hat{a}^{\mathrm{R}}_{\omega \vec{k}_\perp})^\dagger v_{\omega \vec{k}_\perp} (\tau/(a z_n), x_n, y_n, z_n)^* ] \ket{0}_{\mathrm{F}}.
\label{state_field_2}
\end{align}

Considering the expression of the modes in~(\ref{rindler_modes}), we can already compute the integral in~$\tau$, obtaining
\begin{align}
\ket{\omega_i, n}_{\mathrm{F}} =\ & \int_0^\infty \rmd \omega \int \rmd^2 \vec{k}_\perp \sqrt{\frac{\sinh(\pi \omega / a)}{2 \pi^3 a}} K_{\rmi \omega/a} (k_\perp z_n)
\nonumber \\
& \times [\rme^{\rmi \vec{k}_\perp \cdot \vec{x}^\perp_n} \bar{\chi} (\omega_i - \omega/(a z_n)) \hat{a}^{\mathrm{R}}_{\omega \vec{k}_\perp}
\nonumber \\
& + \rme^{-\rmi \vec{k}_\perp \cdot \vec{x}^\perp_n} \bar{\chi} (\omega_i + \omega/(a z_n)) (\hat{a}^{\mathrm{R}}_{\omega \vec{k}_\perp})^\dagger] \ket{0}_{\mathrm{F}};
\label{state_field_3}
\end{align}
where
\begin{equation}
\bar{\chi} (\Omega) := \frac{1}{\sqrt{2 \pi}} \int_{-\infty}^\infty \rmd \tau \chi (\tau) \rme^{\rmi \Omega \tau} = \left( \frac{2}{\pi} \right)^{\frac{1}{4}} T \rme^{-\Omega^2 T^2}
\label{fourier_transform}
\end{equation}
is the Fourier transform of the switching function.

In order to compute now the action of the Rindler annihilation and creation operators $\hat{a}^{\mathrm{R}}_{\omega \vec{k}_\perp}$, $(\hat{a}^{\mathrm{R}}_{\omega \vec{k}_\perp})^\dagger$ on the Minkowski vacuum state, it is convenient to write them in terms of annihilation and creation operators associated to Minkowski modes with well-defined momentum, $\hat{a}^{\mathrm{M}}_{k_z \vec{k}_\perp}$, $(\hat{a}^{\mathrm{M}}_{k_z \vec{k}_\perp})^\dagger$, through a Bogoliubov transformation:
\begin{multline}
\hat{a}^{\mathrm{R}}_{\omega \vec{k}_\perp} = \\
\int_{-\infty}^\infty \rmd k_z [ (\alpha_{\omega k_z k_\perp})^* \hat{a}^{\mathrm{M}}_{k_z \vec{k}_\perp} - (\beta_{\omega k_z k_\perp})^* (\hat{a}^{\mathrm{M}}_{k_z (-\vec{k}_\perp)})^\dagger ],
\label{bogoliubov}
\end{multline}
where $\alpha_{\omega k_z k_\perp}$, $\beta_{\omega k_z k_\perp}$ are the Bogoliubov coefficients between the Minkowski modes and the Rindler modes, given by~\cite{RevModPhys.80.787}
\begin{align}
\alpha_{\omega k_z k_\perp} & = \frac{\rme^{\frac{\omega}{a}\left[\frac{\pi}{2} - \rmi \vartheta(k_z, k_\perp)\right]}}{\sqrt{4 \pi \sqrt{k_z^2 + k_\perp^2} a \sinh(\pi \omega / a)}},
\label{alpha} \\
\beta_{\omega k_z k_\perp} & = -\frac{\rme^{\frac{\omega}{a}\left[-\frac{\pi}{2} - \rmi \vartheta(k_z, k_\perp)\right]}}{\sqrt{4 \pi \sqrt{k_z^2 + k_\perp^2} a \sinh(\pi \omega / a)}};
\label{beta}
\end{align}
with
\begin{equation}
\vartheta(k_z, k_\perp) := \frac{1}{2} \log \left( \frac{\sqrt{k_z^2 + k_\perp^2} + k_z}{\sqrt{k_z^2 + k_\perp^2} - k_z} \right).
\label{rapidity}
\end{equation}

Replacing~(\ref{bogoliubov}), (\ref{alpha}) and~(\ref{beta}) in~(\ref{state_field_3}), we finally get that the state of the field reads
\begin{widetext}
\begin{multline}
\ket{\omega_i, n}_{\mathrm{F}} = \int_0^\infty \rmd \omega \int_{-\infty}^\infty \rmd k_z \int \rmd^2 \vec{k}_\perp \frac{K_{\rmi \omega/a} (k_\perp z_n)}{\sqrt{8 \pi^4 a^2 \sqrt{k_z^2 + k_\perp^2}}}\ \rme^{-\rmi \vec{k}_\perp \cdot \vec{x}^\perp_n} \\
\times \left\{ \rme^{\frac{\omega}{a}\left[-\frac{\pi}{2} + \rmi \vartheta(k_z, k_\perp)\right]} \bar{\chi} (\omega_i - \omega/(a z_n)) + \rme^{\frac{\omega}{a}\left[\frac{\pi}{2} - \rmi \vartheta(k_z, k_\perp)\right]} \bar{\chi} (\omega_i + \omega/(a z_n)) \right\} \ket{\vec{k}_\perp, k_z}_{\mathrm{F}};
\label{state_field_4}
\end{multline}
\end{widetext}
where~$\ket{\vec{k}_\perp, k_z}_{\mathrm{F}}$ is a state with one Minkowski particle with momentum~$(\vec{k}_\perp, k_z)$, with the normalization \footnote{This normalization is not Lorentz invariant~\cite{peskin1995introduction}, but we are going to consider only one Minkowski reference frame~$(T,X,Y,Z)$, so this causes no problems.}
\begin{align}
\ket{\vec{k}_\perp, k_z}_{\mathrm{F}} :=\ & (\hat{a}^{\mathrm{M}}_{k_z \vec{k}_\perp})^\dagger \ket{0}_{\mathrm{F}},
\nonumber \\
\braket{\vec{k}_\perp, k_z|\vec{k}_\perp', k_z'}_{\mathrm{F}} =\ & [\hat{a}_{k_z \vec{k}_\perp}, \hat{a}_{k_z' \vec{k}_\perp'}^\dagger]
\nonumber \\
=\ & \delta(k_z - k_z') \delta^2(\vec{k}_\perp - \vec{k}_\perp').
\label{normalization}
\end{align}

We can see that the state of the field, expanded in the Fock basis associated to Minkowski modes, corresponds to a one-particle state with a certain characteristic dispersion in momentum. This means that, as described by inertial observers, the excitation of the detector is accompanied by the emission of a particle. We can reproduce the state of the field given in~\cite{PhysRevD.38.1118} by taking the limit~$T \to \infty$ in~(\ref{state_field_4}), but the state obtained is not normalizable, as one can easily check \footnote{From the Rindler perspective, the state would correspond to that in which a Rindler particle with well-defined frequency has been absorbed, and since the spectrum is continuous, the norm would be proportional to~$\delta(\omega - \omega)$.}, and therefore not useful for the purposes of computing the distinguishability between different states.

Having already computed the states of the field, it is time now to compute their scalar product. Using~(\ref{state_field_4}) we can write
\begin{widetext}
\begin{align}
\braket{\omega_i, n|\omega_j, m}_{\mathrm{F}} =\ & \int_0^\infty \rmd \omega \int_0^\infty \rmd \omega' \int_{-\infty}^\infty \rmd k_z \int_0^\infty \rmd k_\perp k_\perp \frac{J_0 (k_\perp \Delta x^\perp_{m n}) K_{\rmi \omega/a} (k_\perp z_m) K_{\rmi \omega'/a} (k_\perp z_n)}{4 \pi^3 a^2 \sqrt{k_z^2 + k_\perp^2}}
\nonumber \\
& \times \left[ \rme^{-\frac{\pi}{2a} (\omega + \omega')} \rme^{\frac{\rmi}{a} (\omega - \omega') \vartheta(k_z, k_\perp)} \bar{\chi} (\omega_j - \omega/(a z_m)) \bar{\chi} (\omega_i - \omega'/(a z_n)) \right.
\nonumber \\
& + \rme^{-\frac{\pi}{2a} (\omega - \omega')} \rme^{\frac{\rmi}{a} (\omega + \omega') \vartheta(k_z, k_\perp)} \bar{\chi} (\omega_j - \omega/(a z_m)) \bar{\chi} (\omega_i + \omega'/(a z_n))
\nonumber \\
& + \rme^{\frac{\pi}{2a} (\omega - \omega')} \rme^{-\frac{\rmi}{a} (\omega + \omega') \vartheta(k_z, k_\perp)} \bar{\chi} (\omega_j + \omega/(a z_m)) \bar{\chi} (\omega_i - \omega'/(a z_n))
\nonumber \\
& \left. +\ \rme^{\frac{\pi}{2a} (\omega + \omega')} \rme^{-\frac{\rmi}{a} (\omega - \omega') \vartheta(k_z, k_\perp)} \bar{\chi} (\omega_j + \omega/(a z_m)) \bar{\chi} (\omega_i + \omega'/(a z_n)) \right],
\label{scalar_product}
\end{align}
\end{widetext}
where~$J_\nu (x)$ is the Bessel function of the first kind. We have used~(\ref{normalization}) to trivially evaluate three of the integrals in the momentum, while the angular integral in~$\vec{k}_\perp$ yielded~$~2 \pi J_0 (k_\perp \Delta x^\perp_{m n})$. The integral in~$k_z$ can be also evaluated to
\begin{equation}
\int_{-\infty}^\infty \frac{\rmd k_z}{\sqrt{k_z^2+k_\perp^2}} \rme^{\frac{\rmi}{a} \Omega \vartheta(k_z, k_\perp)} = 2 \pi a \delta(\Omega).
\label{integral_rapidity}
\end{equation}
The result obtained allows then to evaluate the integral in~$\omega'$, for which only two terms in~(\ref{scalar_product}) give a non-zero contribution, resulting in
\begin{widetext}
\begin{align}
\braket{\omega_i, n|\omega_j, m}_{\mathrm{F}} =\ & \int_0^\infty \rmd \omega \int_0^\infty \rmd k_\perp k_\perp \frac{J_0 (k_\perp \Delta x^\perp_{m n}) K_{\rmi \omega/a} (k_\perp z_m) K_{\rmi \omega/a} (k_\perp z_n)}{2 \pi^2 a}\nonumber \\
& \times \left[ \rme^{-\frac{\pi \omega}{a}} \bar{\chi} (\omega_j - \omega/(a z_m)) \bar{\chi} (\omega_i - \omega/(a z_n)) + \rme^{\frac{\pi \omega}{a}} \bar{\chi} (\omega_j + \omega/(a z_m)) \bar{\chi} (\omega_i + \omega/(a z_n)) \right].
\label{scalar_product_2}
\end{align}

At this point, we need to compute an approximation for the integral in~$\omega$, which we take by considering the large time regime given by~(\ref{duration}). In order to do so, we first replace the explicit form of~$\bar{\chi} (\Omega)$ in~(\ref{fourier_transform}), obtaining after some manipulation
\begin{align}
\braket{\omega_i, n|\omega_j, m}_{\mathrm{F}} =\ & \frac{T^2 \rme^{-C}}{\sqrt{2 \pi^5} a} \int_0^\infty \rmd k_\perp k_\perp J_0 (k_\perp \Delta x^\perp_{m n}) \int_0^\infty \rmd \omega\ K_{\rmi \omega/a} (k_\perp z_m) K_{\rmi \omega/a} (k_\perp z_n)
\nonumber \\
& \times \left[ \rme^{-\frac{\pi \omega}{a}} \rme^{-(\omega/\bar{\omega} - 1)^2 M} + \rme^{\frac{\pi \omega}{a}} \rme^{-(\omega/\bar{\omega} + 1)^2 M} \right];
\label{replace_chi}
\end{align}
\end{widetext}
where
\begin{align}
C & := \frac{(\omega_j z_m - \omega_i z_n)^2}{z_m^2 + z_n^2}\ T^2, \quad \bar{\omega} := a\ \frac{\omega_j / z_m + \omega_i / z_n}{1/z_m^2 + 1/z_n^2},
\nonumber \\
M & := \frac{(\omega_i z_m + \omega_j z_n)^2}{z_m^2 + z_n^2}\ T^2.
\label{quantities_gaussians}
\end{align}

The two terms obtained are Gaussian functions of width~$\bar{\omega}/\sqrt{2M}$. The second term is peaked at the negative value~$-\bar{\omega}$. Noticing that, because of~(\ref{duration}), $\sqrt{2M} \gtrsim 2/\varepsilon \gg 1$, we have that~$|- \bar{\omega}| \gg \bar{\omega}/\sqrt{2M}$, and therefore the contribution of the second term to the integral in positive~$\omega$ is negligible. Also because~$M \gg 1$, we can use Laplace's method to approximate the integral of the first term with high accuracy (with a \emph{relative} error of~$O(1/M) \sim O(\varepsilon^2)$). Before doing so, let us however center our attention on the factor~$\rme^{-C}$. This factor will be very close to zero unless
\begin{equation}
\frac{|\omega_j z_m - \omega_i z_n|}{\sqrt{z_m^2 + z_n^2}}\ T \lesssim \frac{1}{\sqrt{2}}.
\label{strict_approximation}
\end{equation}
Notice that, when considering the characteristic range of frequencies \footnote{The expected result from what is known from the standard Unruh effect is that the characteristic range of frequencies involved is of the order of the acceleration (in natural units). This will be also the case for the results obtained in this work.} $\omega_i \sim 1/z_n$ and $\omega_j \sim 1/z_m$, this condition can be written as
\begin{equation}
\left| \omega_j z_m - \omega_i z_n \right| \lesssim \varepsilon.
\label{approximation_quotients}
\end{equation}
Although for high frequencies in the spectrum this condition might be more restrictive than~(\ref{strict_approximation}), it is in any case a \emph{sufficient} condition, and also \emph{necessary} within the characteristic range of frequencies. Therefore, for simplicity we will assume it as the condition for the factor~$\rme^{-C}$ not to become negligible. We conclude then that the scalar product that we are computing is only significant if the relation between quotients
\begin{equation}
\omega_i z_n \approx \omega_j z_m
\label{quotients}
\end{equation}
holds, with the limit for the validity of the approximation given by~(\ref{approximation_quotients}). This is precisely the condition in~(\ref{quotients_q}).

Since we assume the approximation~(\ref{quotients}) to be accurate to first order in~$\varepsilon$, when it holds it is legitimate to use relation~(\ref{quotients}) in the calculations. In such case, we have that~$C \approx 0$, $\bar{\omega} \approx a \omega_j z_m$ and~$M \approx (1 + z_m^2/z_n^2) (\omega_j T)^2 \gg 1$. With this, we proceed to approximate the integral in~$\omega$ of the first term in~(\ref{replace_chi}) using Laplace's method, obtaining
\begin{multline}
\braket{\omega_i, a_n|\omega_j, a_m}_{\mathrm{F}} \approx\ \frac{T \rme^{-\pi \omega_j z_m}}{\sqrt{2 \pi^4 (1/z_m^2 + 1/z_n^2)}} \int_0^\infty \rmd k_\perp k_\perp
\\
\times J_0 (k_\perp \Delta x^\perp_{m n}) K_{\rmi \omega_j z_m} (k_\perp z_m) K_{\rmi \omega_j z_m} (k_\perp z_n).
\label{large_time}
\end{multline}

In the following, abusing notation we will consider the approximate expressions obtained using~(\ref{duration}) as exact, so that the value of the scalar product~$\braket{\omega_i, n|\omega_j, m}_{\mathrm{F}}$ is directly given by~(\ref{large_time}) when~(\ref{quotients}) is fulfilled, and vanishes in other case.

The remaining integral in~$k_\perp$ in~(\ref{large_time}) has to be computed separately for the case in which the trajectories are the same, $n = m$ (and thus~$\omega_i = \omega_j$ within the approximation), and for the case in which they are different. In the first case, we obtain
\begin{equation}
\braket{\omega_j, m|\omega_j, m}_{\mathrm{F}} = \frac{T}{2 \pi} \frac{\omega_j}{\rme^{2 \pi \omega_j z_m} - 1}.
\label{scalar_product_diagonal}
\end{equation}
With this result we have already obtained the diagonal terms in~(\ref{state_detector}). In order to compute the off-diagonal terms, it is more convenient to compute directly the scalar product between the normalized states of the field~$\Lambda^{ij}_{nm}$ in~(\ref{lambda_def}). Using~(\ref{large_time}) and~(\ref{scalar_product_diagonal}), and implementing the variable transformation~$k_\perp = \bar{k} \sqrt{1/z_m^2+1/z_n^2}$ in the integral in~(\ref{large_time}), we obtain
\begin{widetext}
\begin{equation}
\Lambda^{ij}_{nm} = \frac{2 \sinh (\pi q_{j m})}{\pi q_{j m}} \sqrt{\cosh \Delta \xi_{m n}} \int_0^\infty \rmd \bar{k}\ \bar{k}\ J_0 (\bar{k} \Delta \bar{x}_{m n}) K_{\rmi q_{j m}} \left(\bar{k} \sqrt{\frac{\rme^{2 \Delta \xi_{m n}} + 1}{2}} \right) K_{\rmi q_{j m}} \left(\bar{k} \sqrt{\frac{\rme^{-2 \Delta \xi_{m n}} + 1}{2}} \right),
\label{overlapping}
\end{equation}
\end{widetext}
with the definition of the quantities~$q_{j m}$, $\Delta \xi_{m n}$ and~$\Delta \bar{x}_{m n}$ given in~(\ref{q_def}) and~(\ref{relative_values}). If we solve the integral in~$\bar{k}$ (see Eq.~6.578.10 in~\cite{gradshtein1980table} for the analytic solution), we get the expression for~$\Lambda^{ij}_{nm}$ in~(\ref{f_weight}). Trivially solving for~$\braket{\omega_i, n|\omega_j, m}_{\mathrm{F}}$ in~(\ref{lambda_def}) and using again~(\ref{scalar_product_diagonal}) gives the coefficients of the off-diagonal terms in~(\ref{state_detector}), completing the proof.

We highlight again the way we have proceed to compute the normalized scalar product in~(\ref{overlapping}). If we follow the computations, we realize that this quantity is obtained by \emph{first} computing the scalar product between different states for fixed~$T$, and \emph{then} taking the large-time limit of this scalar product in~(\ref{large_time}). The result of the limit computed is the correct description of any physically realistic scenario, in which the interaction time can be in principle arbitrarily large but finite. Trying to take first the large time limit directly in the expression for the state~(\ref{state_field_4}) leads, as we already mentioned, to non-normalizable states, which degree of distinguishability is not defined.

We can at this point justify the need for the limitation~(\ref{acceleration_limitation}). Indeed, we need the perturbative contributions in~(\ref{state_detector}) to remain~$O(\varepsilon)$ or smaller. Since being a scalar product we have that $|\Lambda^{ij}_{nm}| \leq 1$, the off-diagonal contributions are always smaller than the diagonal ones in~(\ref{state_detector_gen}), and we just need to check the order of magnitude of these. They reach their maximum for the lowest frequency~$\omega_1$ and the highest acceleration~$a_1 = 1/z_1$, in which case using~(\ref{duration}) we have that
\begin{equation}
\braket{\omega_1, 1|\omega_1, 1}_{\mathrm{F}} \sim \frac{1}{2 \pi \varepsilon (\rme^{2 \pi \omega_1 / a_1} - 1)} \lesssim \frac{1}{\varepsilon} \quad \Leftrightarrow \quad \frac{\omega_1}{a_1} \gtrsim \mu,
\label{limit_scalar_products}
\end{equation}
which is condition~(\ref{acceleration_limitation}).

Finally, we also notice that, if we rescale all~$\omega_i \to \gamma \omega_i$, all~$z_n \to z_n/\gamma$, and~$T \to T/\gamma$, with~$\gamma > 0$, all results still hold and remain identical. This is consistent with the fact that a massless field does not introduce any privileged scale.

%-----------------------------------------------------------------------------------------------
\section{Scalar product~$\Lambda^{ij}_{nm}$. Further expressions}\label{function_f}
%-----------------------------------------------------------------------------------------------

\begin{figure}[h]
\includegraphics[width=\columnwidth]{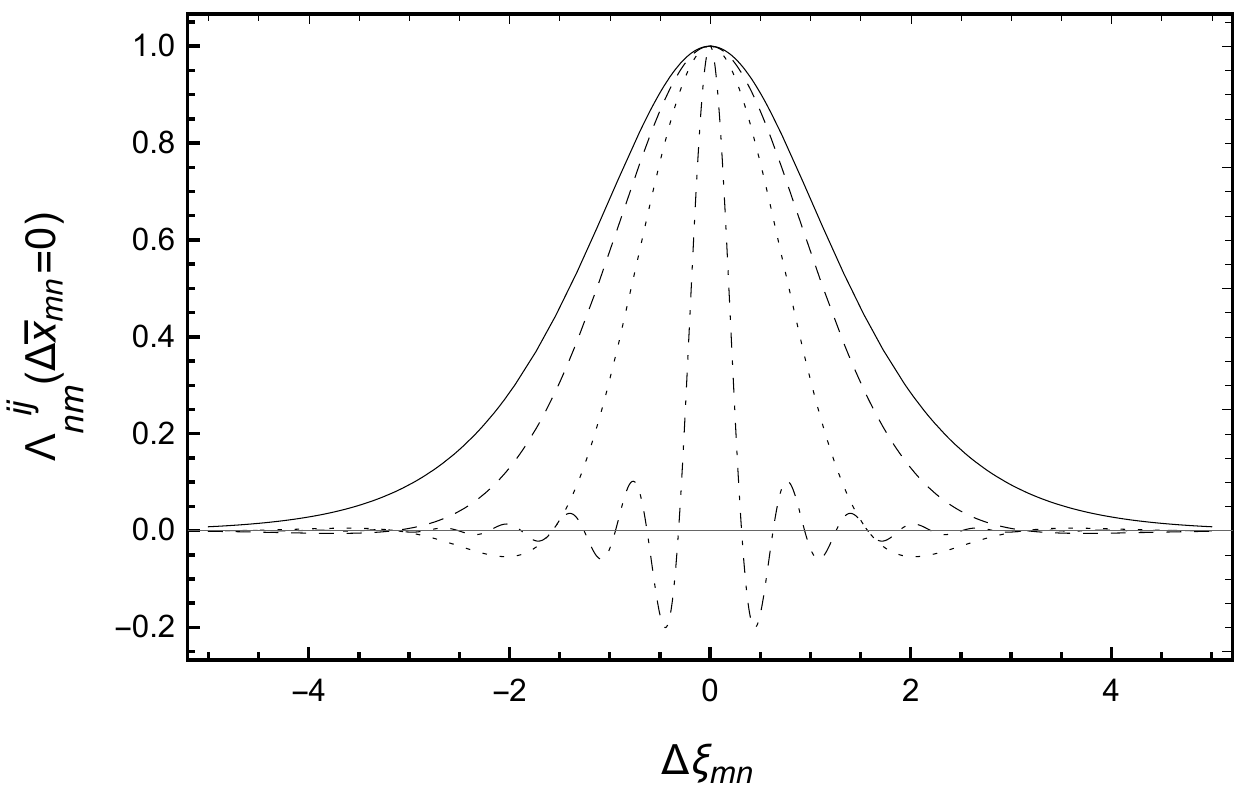}
\caption{\label{overlapping_xi} Scalar product~$\Lambda^{ij}_{nm}(\Delta \bar{x}_{mn} = 0)$ as a function of~$\Delta \xi_{m n}$ for~$q_{j m} = 0$ (solid), $q_{j m} = 1$ (dashed), $q_{j m} = 2$ (dotted) and $q_{j m} = 10$ (dash-dot).}
\end{figure}

\begin{figure}[h]
\includegraphics[width=\columnwidth]{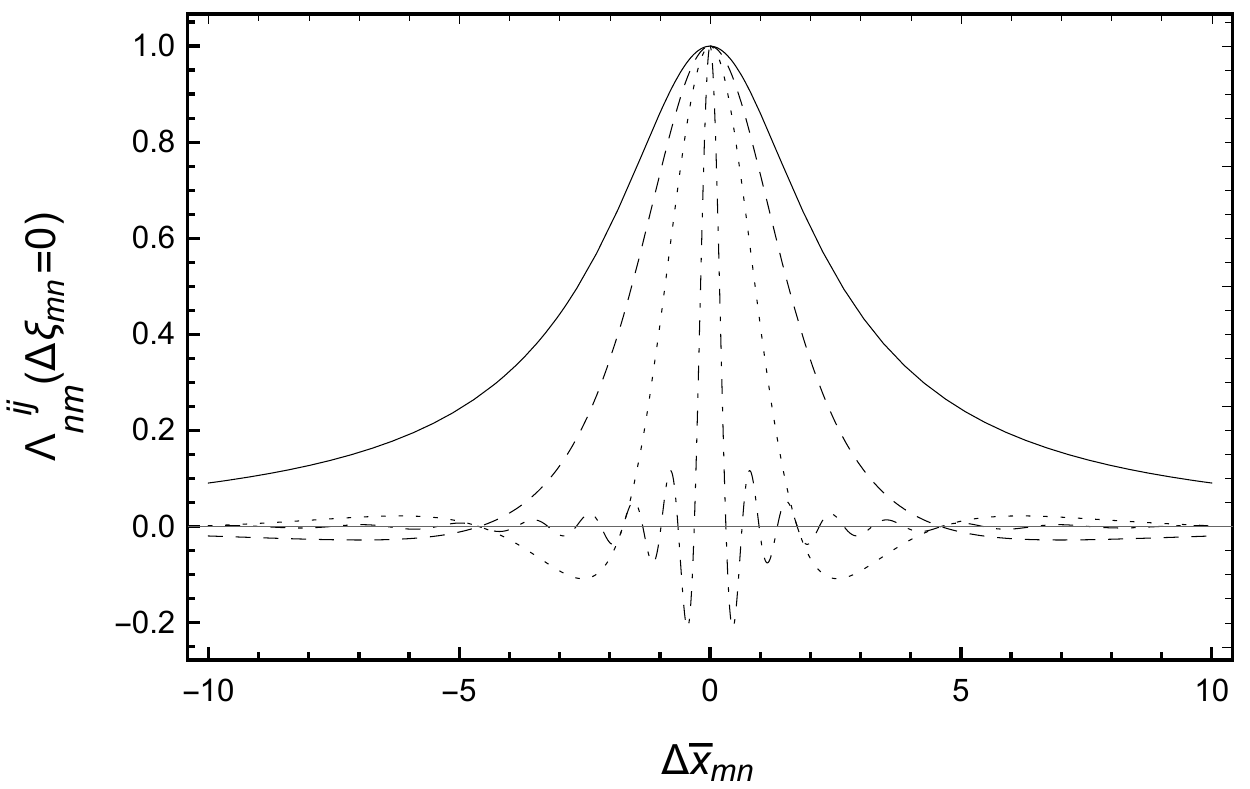}
\caption{\label{overlapping_x} Scalar product~$\Lambda^{ij}_{nm}(\Delta \xi_{mn} = 0)$ as a function of~$\Delta \bar{x}_{m n}$ for~$q_{j m} = 0$ (solid), $q_{j m} = 1$ (dashed), $q_{j m} = 2$ (dotted) and $q_{j m} = 10$ (dash-dot).}
\end{figure}

The scalar product~$\Lambda^{ij}_{nm}$ in~(\ref{f_weight}) has very simple analytic expressions for the cases~$\Delta \bar{x}_{mn} = 0$ and~$\Delta \xi_{mn} = 0$. These expressions are
\begin{align}
\Lambda^{ij}_{nm}(\Delta \bar{x}_{mn} = 0) & = \frac{\sin (q_{j m} \Delta \xi_{mn})}{q_{j m}} \frac{\csch \Delta \xi_{mn}}{\sqrt{\cosh \Delta \xi_{mn}}},
\label{f_xi} \\
\Lambda^{ij}_{nm}(\Delta \xi_{mn} = 0) & = \frac{\sin (q_{j m} g (\Delta \bar{x}_{mn}))}{q_{j m}} \csch g (\Delta \bar{x}_{mn});
\label{f_x}
\end{align}
with
\begin{equation}
g (\Delta \bar{x}_{mn}) := 2 \arcsinh \left( \frac{\Delta \bar{x}_{mn}}{2} \right).
\label{def_g}
\end{equation}
In Figures~\ref{overlapping_xi} and~\ref{overlapping_x} we plot these functions for different values of~$q_{j m}$.


%-----------------------------------------------------------------------------------------------
\section{Continuous degrees of freedom}\label{continuous}
%-----------------------------------------------------------------------------------------------

We briefly consider the case in which the detector has a continuous spectrum and is spread in position continuously in Rindler coordinates \footnote{One could also consider taking to the continuum only one of the degrees of freedom, either the external or the internal.}.

While taking the position to the continuum is relatively trivial and does not introduce much novelty, considering a continuous spectrum allows to strictly take the limit of the duration of the interaction going to infinity~$T \to \infty$ without obtaining diverging probabilities (since what we are interested now is probability \emph{densities}), and also to simplify the formal result, avoiding the necessity of ``conditional sums'' as in~(\ref{state_detector}). We will use the computations done for the discrete case when they also hold for the continuous case, pointing out just the expressions that have to be changed. We also assume that the meaning of the new notation for the continuous case can be inferred without explicit clarifications.

First, we replace the monopole moment of the detector in~(\ref{monopole}) by
\begin{equation}
\hat{m} (\tau) = \int_0^\infty \rmd \omega \int_0^\infty \rmd \omega' \zeta (\omega, \omega') \rme^{\rmi (\omega - \omega') \tau} \ket{\omega} \bra{\omega'}_{\mathrm{D}},
\label{monopole_cont}
\end{equation}
where~$\zeta (\omega', \omega) = \zeta (\omega, \omega')^*$ and the internal levels are normalized to~$\braket{\omega' | \omega} = \delta(\omega' - \omega)$. In the following calculations we will use the simplification $\zeta(\omega) \equiv \zeta(\omega, 0)$, since only this quantity will appear. The initial state of the system in~(\ref{initial_state}) changes to
\begin{equation}
\ket{\Psi (\tau \to -\infty)} = \ket{0}_{\mathrm{D}} \ket{0}_{\mathrm{F}} \int \rmd^3 \vec{x} A(\vec{x}) \ket{\vec{x}}_{\mathrm{T}},
\label{initial_state_cont}
\end{equation}
with~$\braket{\vec{x}' | \vec{x}} = \delta^3 (\vec{x}' - \vec{x})$.

The final state in~(\ref{final_state_2}) reads now
\begin{multline}
\ket{\Psi (\tau \to \infty)} \approx \ket{0}_{\mathrm{D}} \ket{0}_{\mathrm{F}} \int \rmd^3 \vec{x} A(\vec{x}) \ket{\vec{x}}_{\mathrm{T}} \\
+ \rmi \varepsilon \int_0^\infty \rmd \omega \int \rmd^3 \vec{x} \zeta(\omega) A(\vec{x}) \ket{\omega}_{\mathrm{D}} \ket{\omega, \vec{x}}_{\mathrm{F}} \ket{\vec{x}}_{\mathrm{T}},
\label{final_state_2_cont}
\end{multline}
and the result of tracing out the field in~(\ref{state_detector_gen}) changes to
\begin{multline}
\rho_{\mathrm{DT}} \approx \ket{0}\bra{0}_{\mathrm{D}} \left( \int \rmd^3 \vec{x} \int \rmd^3 \vec{x}' A(\vec{x}')^* A(\vec{x}) \ket{\vec{x}}\bra{\vec{x}'}_{\mathrm{T}} \right) \\
+ \varepsilon^2 \int \rmd^3 \vec{x} \int \rmd^3 \vec{x}' \int_0^\infty \rmd \omega \int_0^\infty \rmd \omega' \zeta(\omega')^* A(\vec{x}')^* \zeta(\omega) A(\vec{x}) \\
\times \braket{\omega', \vec{x}'|\omega, \vec{x}}_{\mathrm{F}} \ket{\omega}\bra{\omega'}_{\mathrm{D}} \ket{\vec{x}}\bra{\vec{x}'}_{\mathrm{T}}.
\label{state_detector_gen_cont}
\end{multline}
The computation of the scalar products~$\braket{\omega', \vec{x}'|\omega, \vec{x}}_{\mathrm{F}}$ in Appendix~\ref{computation} follows in an identical way until taking the large time limit. The expression right before taking this limit~(\ref{replace_chi}), obtained after replacing the switching functions, is
\begin{multline}
\braket{\omega', \vec{x}'|\omega, \vec{x}}_{\mathrm{F}} = \frac{T^2 \rme^{-C}}{\sqrt{2 \pi^5} a} \int_0^\infty \rmd k_\perp k_\perp J_0 (k_\perp \Delta x^\perp) \\
\times \int_0^\infty \rmd \omega'' K_{\rmi \omega''/a} (k_\perp z) K_{\rmi \omega''/a} (k_\perp z') \\
\times \left[ \rme^{-\frac{\pi \omega''}{a}} \rme^{-(\omega''/\bar{\omega} - 1)^2 M} + \rme^{\frac{\pi \omega''}{a}} \rme^{-(\omega''/\bar{\omega} + 1)^2 M} \right];
\label{replace_chi_cont}
\end{multline}
with the quantities~$C$, $\bar{\omega}$ and~$M$ given by~(\ref{quantities_gaussians}) with the corresponding notation replacements.

It is easy to notice that, due to the factor~$\rme^{-C}$, when taking the limit~$T \to \infty$ the scalar product~(\ref{replace_chi_cont}) vanishes unless~$\omega' z' = \omega z$, in which case it diverges. Notice that, since we are taking the strict limit, unlike in~(\ref{quotients}) now the relation has to hold exactly. Since we have a function that vanishes everywhere except on a point where it diverges, checking that its integral remains finite when taking the limit~$T \to \infty$ suffices to prove that in this limit we have a Dirac delta. We take then the following integral:
\begin{multline}
\int \rmd \omega' \braket{\omega', \vec{x}'|\omega, \vec{x}}_{\mathrm{F}} = \frac{T}{\sqrt{2 \pi^4} a} \int_0^\infty \rmd k_\perp k_\perp J_0 (k_\perp \Delta x^\perp) \\
\times \int_0^\infty \rmd \omega'' K_{\rmi \omega''/a} (k_\perp z) K_{\rmi \omega''/a} (k_\perp z') \\
\times \left\{ \rme^{-\frac{\pi \omega''}{a}} \rme^{-[\omega''/(a \omega z) - 1]^2 \omega^2 T^2} \right. \\
\left. + \rme^{\frac{\pi \omega}{a}} \rme^{-[\omega''/(a \omega z) + 1]^2 \omega^2 T^2} \right\}.
\label{integral_omegap_cont}
\end{multline}
In the limit~$T \to \infty$, the second term does not contribute to the integral in~$\omega''$, while the first term is given exactly by Laplace's method. We obtain
\begin{multline}
\int \rmd \omega' \braket{\omega', \vec{x}'|\omega, \vec{x}}_{\mathrm{F}} = \\
\frac{z \rme^{- \pi q}}{\sqrt{2 \pi^3}} \int_0^\infty \rmd k_\perp k_\perp J_0 (k_\perp \Delta x^\perp) K_{\rmi q} (k_\perp z) K_{\rmi q} (k_\perp z') = \\
\frac{\sqrt{\cosh \Delta \xi}}{z' \sqrt{2 \pi}} \Lambda(q, \Delta \xi, \Delta \bar{x}) \frac{q}{\rme^{2 \pi q}-1},
\label{integral_omegapp_cont}
\end{multline}
where the different quantities are defined in~(\ref{q_def}), (\ref{relative_values}) and~(\ref{f_weight}), with the obvious change of notation. Summarizing, we can write that
\begin{multline}
\braket{\omega', \vec{x}'|\omega, \vec{x}}_{\mathrm{F}} = \\
\frac{\sqrt{\cosh \Delta \xi}}{z' \sqrt{2 \pi}} \Lambda(q, \Delta \xi, \Delta \bar{x}) \frac{q}{\rme^{2 \pi q}-1} \delta (\omega' - \omega z/z').
\label{delta_omegap}
\end{multline}

The Dirac delta obtained allows us to compute the integral in~$\omega'$ in~(\ref{state_detector_gen_cont}), and we finally obtain, after a trivial change of variable~$\omega = q / z$,
\begin{align}
\rho_{\mathrm{DT}} =\ & \ket{0}\bra{0}_{\mathrm{D}} \left( \int \rmd^3 \vec{x} \int \rmd^3 \vec{x}' A(\vec{x}')^* A(\vec{x}) \ket{\vec{x}}\bra{\vec{x}'}_{\mathrm{T}} \right)
\nonumber \\
& + \frac{\varepsilon^2}{\sqrt{2 \pi}} \int \rmd^3 \vec{x} \int \rmd^3 \vec{x}' A(\vec{x}')^* A(\vec{x}) \sqrt{\frac{1/z^2+1/z'^2}{2}}
\nonumber \\
& \times \int_0^\infty \rmd q\ \zeta(q/z')^* \zeta(q/z) \Lambda(q, \Delta \xi, \Delta \bar{x})
\nonumber \\
& \times \frac{q/\sqrt{z z'}}{\rme^{2 \pi q}-1} \ket{q/z}\bra{q/z'}_{\mathrm{D}} \ket{\vec{x}}\bra{\vec{x}'}_{\mathrm{T}}.
\label{state_detector_cont}
\end{align}

We observe again the planckian spectrum in the diagonal terms, and this same spectrum weighted by the function~$\Lambda$ in the off-diagonal terms. The different factors appearing as compared to~(\ref{state_detector}) respond to the slightly different construction required in the continuous case. Notice that recovering the results for the discrete case out of~(\ref{state_detector_cont}) is not trivial, since unlike in~(\ref{state_detector}) the strict limit~$T \to \infty$ has been taken. In particular, trying to simply use some Dirac comb-like distribution for~$\zeta(\omega)$ clearly yields diverging results.

\begin{acknowledgments}
The authors want to thank Flaminia Giacomini, Philippe A.\ Gu\'erin, Ilya Kull, Aleksandra Dimic, Dragoljub Gocanin, Marko Milivojevic and Magdalena Zych for useful discussions during the development of this work, and specially Jorma Louko for pointing out the analytic solution for~(\ref{f_weight}). L.~C.~B.\ acknowledges the support from the research platform TURIS and from the European Commission via Testing the Large-Scale Limit of Quantum Mechanics (TEQ) (No.~766900) project. E.~C-R.\ is supported in part by the Program of Concerted Research Actions (ARC) of the Universit\'e Libre de Bruxelles. The authors were also supported by the Austrian-Serbian bilateral scientific cooperation no.\ 451-03-02141/2017-09/02, and by the Austrian ScienceFund (FWF) through the SFB project ``BeyondC'' and a grant from the Foundational Questions Institute (FQXi) Fund. This publication was made possible through the support of the ID\#~61466 grant from the John Templeton Foundation, as part of the ``The Quantum Information Structure of Spacetime (QISS)'' Project (qiss.fr). The opinions expressed in this publication are those of the author(s) and do not necessarily reflect the views of the John Templeton Foundation.
\end{acknowledgments}

\bibliography{biblio}

\end{document}